\documentclass[prc,showpacs,preprintnumbers,amsmath,amssymb,superscriptaddress,groupedaddress,nofootinbib,tightenlines]{revtex4}

\usepackage{amssymb}
\usepackage{graphicx}
\usepackage{amsmath}
\usepackage{bm}
\usepackage{bbm}
\usepackage{color}
\usepackage{hyperref}
\usepackage{slashed}

\def\beqn{\begin{eqnarray}}
\def\eeqn{\end{eqnarray}}
\def\barr{\begin{array}}
\def\earr{\end{array}}
\def\btab{\begin{tabular}}
\def\etab{\end{tabular}}
\def\bite{\begin{itemize}}
\def\eite{\end{itemize}}
\def\bcen{\begin{center}}
\def\ecen{\end{center}}

\def\eq{\begin{equation}}
\def\ee{\end{equation}}

\def\q2dagger{q_2\hspace{-0.35cm}/\;}

\newcommand{\bea}{\begin{eqnarray}}
\newcommand{\eea}{\end{eqnarray}}

\begin{document}
\title{Dispersion relation analysis of the radiative corrections to $g_A$ in the neutron $\beta$-decay}
\author{Mikhail Gorchtein}
\affiliation{Institut f\"ur Kernphysik, PRISMA Cluster of Excellence\\
Johannes Gutenberg-Universit\"at, Mainz, Germany}
\email{gorshtey@uni-mainz.de}
\author{Chien-Yeah Seng}
\affiliation{Helmholtz-Institut f\"{u}r Strahlen- und Kernphysik and Bethe Center for
  Theoretical Physics,\\ Universit\"{a}t Bonn, 53115 Bonn, Germany}

\date{\today}

\begin{abstract}
	
We present the first and complete dispersion relation analysis of the inner radiative corrections to the axial coupling constant $g_A$ in the neutron $\beta$-decay. Using experimental inputs from the elastic form factors and the spin-dependent structure function $g_1$, we determine the contribution from the $\gamma W$-box diagram to a precision better than $10^{-4}$. Our calculation indicates that the inner radiative corrections to the Fermi and the Gamow-Teller matrix element in the neutron $\beta$-decay are almost identical, i.e. the ratio $\lambda=g_A/g_V$ is almost unrenormalized. With this result, we predict the bare axial coupling constant to be {$\mathring{g}_A=-1.2754(13)_\mathrm{exp}(2)_\mathrm{RC}$} based on the PDG average $\lambda=-1.2756(13)$.
\end{abstract}
\pacs{}
\maketitle
\section{Introduction}

\noindent
The recent emergence of an apparent deficit in the top-row Cabibbo-Kobayashi-Maskawa (CKM) matrix 
unitarity~\cite{Zyla:2020zbs},
\begin{equation}
|V_{ud}|^2+|V_{us}|^2+|V_{ub}|^2=0.9985(5)
\end{equation}
has triggered a renewed interest in precise experimental studies of various $\beta$-decay processes giving access to $|V_{ud}|$. Superallowed  $0^+-0^+$ nuclear decays have long been regarded as the best avenue for such purpose. Recent works, however, pointed out that current theory uncertainties in the nuclear-structure corrections may have been significantly underestimated~\cite{Seng:2018yzq,Seng:2018qru,Gorchtein:2018fxl,Hardy:2020qwl}. Reducing these uncertainties  requires novel ab-initio nuclear theory calculations that are not yet available. As a consequence, the role of alternative channels such as the $\beta$-decays of the free neutron, mirror nuclei and pion becomes increasingly important. With the future improvement in the experimental precision, these other $\beta$ decay processes will offer competitive determinations of $|V_{ud}|$ and complementary sensitivity to possible beyond standard model (BSM) signals.

Free neutron $\beta$-decay currently provides the second best determination of $|V_{ud}|$ through the following master formula~\cite{Czarnecki:2004cw,Seng:2020wjq}:
\begin{equation}
|V_{ud}|^2=\frac{4903.1(1.1)~\mathrm{s}}{\tau_n(1+3\lambda^2)}~,
\end{equation}
where the uncertainty in the numerator arises from the Standard Model (SM) theory input. The two required experimental inputs are  
the neutron lifetime $\tau_n$, and the decay parameter $\lambda\equiv g_A/g_V$ which is the ratio between the neutron axial and vector coupling constant. This parameter is renormalized by electroweak radiative corrections (RCs), and these latter are the  primary focus of this article.

The parameter $\lambda$ can be measured either via the P-even correlation $\vec{p}_e\cdot\vec{p}_\nu$ (the $a$ coefficient), or the P-odd $\hat{e}_s\cdot\vec{p}_e$ (the $A$ coefficient) and $\hat{e}_s\cdot\vec{p}_\nu$ (the $B$ coefficient) ones, with $\hat e_s$ the unit 3-vector along the neutron polarization. The current best measurement reported by the PERKEO III collaboration $\lambda=-1.27641(45)_\mathrm{stat}(33)_\mathrm{sys}$, with a 0.04\% precision~\cite{Markisch:2018ndu} (in this paper we pick the sign convention $\lambda <0$, also adopted in the Particle Data Group (PDG) review). However, the current PDG average reads $\lambda=-1.2756(13)$~\cite{Zyla:2020zbs}, where the much larger uncertainty is due to a scale factor of 2.6 that accounts for the large discrepancy between the results before~\cite{Bopp:1986rt,Erozolimsky:1997wi,Liaud:1997vu,Mostovoi:2001ye} and after 2002~\cite{Schumann:2007hz,Mund:2012fq,Darius:2017arh,Brown:2017mhw,Markisch:2018ndu} (see Ref.\cite{Czarnecki:2018okw} for more discussions). Future improvements are expected from the Nab~\cite{Fry:2018kvq} and PERC~\cite{Dubbers:2007st,Wang:2019pts} collaborations, both aiming at an accuracy level of $10^{-4}$. 

The exact value of $\lambda$ not only serves for extracting $V_{ud}$, but is also interesting in itself. 
The ``bare'' (i.e. without electroweak corrections) neutron axial coupling $\mathring{g}_A$ is one of the simplest hadronic matrix elements and has received much attention. Unlike its vector counterpart $\mathring{g}_V$ which remains non-renormalized due to the conserved vector current (CVC), the bare axial coupling is not protected and must be calculated, e.g. using lattice Quantum Chromodynamics (QCD) ~\cite{Khan:2006de,Lin:2008uz,Capitani:2012gj,Horsley:2013ayv,Bali:2014nma,Abdel-Rehim:2015owa,Alexandrou:2017hac,Capitani:2017qpc,Edwards:2005ym,Yamazaki:2008py,Yamazaki:2009zq,Bratt:2010jn,Green:2012ud,Yamanaka:2018uud,Liang:2018pis,Ishikawa:2018rew,Ottnad:2018fri,Bhattacharya:2016zcn,Berkowitz:2017gql,Chang:2018uxx,Gupta:2018qil,Walker-Loud:2019cif}. The most recent FLAG average~\cite{Aoki:2019cca} reads:
\begin{eqnarray}
N_f=2+1+1:&&\mathring{g}_A=-1.251(33)\nonumber\\
N_f=2+1:&&\mathring{g}_A=-1.254(16)(30)\nonumber\\
N_f=2:&&\mathring{g}_A=-1.278(86)~,
\end{eqnarray}
but individual calculations have achieved higher precision. For instance, Ref.\cite{Chang:2018uxx} reported a percent-level determination of $\mathring{g}_A=-1.271(10)(7)$ using an unconventional method inspired by the Feynman-Hellmann theorem, and follow-up works are aiming for sub-percent precision~\cite{Walker-Loud:2019cif}. Such a rapid development makes $\lambda$ a powerful tool for searching for {{red}BSM physics}. By comparing first-principles calculations of $\mathring{g}_A$ to the experimental results for $\lambda$
one thereby constraints the strength of possible BSM contributions that could modify $g_A$, in particular the right-handed currents~\cite{Gonzalez-Alonso:2016etj,Alioli:2017ces,Gonzalez-Alonso:2018omy,Falkowski:2020pma}. 

When the lattice precision reaches $10^{-3}$, 
a valid comparison between $\mathring{g}_A$ and $\lambda$ will require precise $\mathcal{O}(\alpha_{em}/\pi)$ RCs that bring $\mathring{g}_A/\mathring{g}_V$ to $g_A/g_V$. In particular, we need to deal with sizable hadronic uncertainties originating from the $\gamma W$-box diagram. 
The latter can be written as a $Q^2$-integral, and performing operator product expansion (OPE) it is well-known that the corrections to $g_V$ and $g_A$ coming from large $Q^2$ (which carries a large electroweak logarithm) are the same. Therefore, it was believed that the difference between $\mathring{g}_A$ and $\lambda$ is numerically small~\cite{Sirlin:1967zza,Garcia:1983yy,Kurylov:2001av,Kurylov:2002vj}. However, for a long time there was no serious attempt to understand the RC to $g_A$ from the low-$Q^2$ part of the integral which is also of the order $10^{-3}$, comparable to the high-$Q^2$ contribution. It includes the elastic contributions that are fixed by the nucleon form factors, as well as the inelastic contributions that are governed by non-perturbative QCD. The first attempt for a complete analysis was performed recently in Refs.\cite{Hayen:2020cxh,Hayen:2021iga}. Anticipating the results of this work, we found that those Refs. originally contained algebraic mistakes in the computation of the elastic contribution, invalidating their numerical results. {These mistakes were later corrected in the published version of Ref.\cite{Hayen:2020cxh}.} Additionally, the inelastic contribution residing at low $Q^2$ was obtained based on a holographic QCD model, following Ref.\cite{Czarnecki:2019mwq} where the RC to $g_V$ was addressed. The model-dependent nature of this approach makes a rigorous estimation of the theoretical uncertainty complicated.\\

In this article we improve on both points. We perform a novel analysis of the RC to $g_A$ based on the 
dispersion relation (DR) approach. It is a powerful tool which has proved successful in the treatment of the RCs to the Fermi amplitude in the free neutron and superallowed $\beta$-decays~\cite{Seng:2018yzq,Seng:2018qru,Gorchtein:2018fxl,Seng:2020wjq}. In this formalism, the  $\gamma W$-box diagram is expressed as a dispersion integral over structure functions that are directly or indirectly related to experimental data. This ultimately allows for a fully data-driven analysis of this RC. For $g_A$, the required input relies on the spin-dependent structure functions $g_1$ and $g_2$, well-studied quantities in deep inelastic scattering (DIS) experiments. We utilize high-precision world data {on} $g_{1,2}$ to evaluate the dispersion integral, and fix the forward $\gamma W$-box diagram correction to $g_A$ to an unprecedented precision better than $10^{-4}$. We observe that the RCs to $g_V$ and $g_A$ are numerically very close and largely cancel in the ratio, which practically removes any distinction between $\mathring{g}_A$ and $\lambda$ down to $2\times 10^{-4}$. 

The contents in this paper are arranged as follows. In Section~\ref{sec:framework} we define our notation and introduce the starting point for the discussion of the RC. We introduce the $\gamma W$-box diagram in Section~\ref{sec:box}, and derive its dispersive representation in Section~\ref{sec:DR}. The elastic (Born) and inelastic contributions to the box diagram are computed in Section~\ref{sec:born} and \ref{sec:inelastic} respectively. The final results and discussions are presented in Section~\ref{sec:final}.

\section{General framework}
\label{sec:framework}

We start by defining the hadronic currents relevant to the $\beta$-decay of the free neutron:
\begin{eqnarray}
J_{em}^\mu&=&\frac{2}{3}\bar{u}\gamma^\mu u-\frac{1}{3}\bar{d}\gamma^\mu d\nonumber\\
J_W^\mu &=&\bar{u}\gamma^\mu(1-\gamma_5)d~.
\end{eqnarray}
Their single-nucleon matrix elements are given by:
\begin{eqnarray}
\langle N(p_f,s_f)|J_{em}^\mu|N(p_i,s_i)\rangle&=&\bar{u}_{s_f}(p_f)\left[F_1^N\gamma^\mu+\frac{i}{2M}F_2^N\sigma^{\mu\nu}(p_f-p_i)_\nu\right]u_{s_i}(p_i)\nonumber\\
\langle p(p_f,s_f)|J_W^\mu|n(p_i,s_i)\rangle&=&\bar{u}_{s_f}(p_f)\left[F_1^W\gamma^\mu+\frac{i}{2M}F_2^W\sigma^{\mu\nu}(p_f-p_i)_\nu+G_A\gamma^\mu\gamma_5-\frac{G_P}{2M}\gamma_5(p_f-p_i)^\mu\right]u_{s_i}(p_i)~,\label{eq:formfactors}
\end{eqnarray}
where $N=p,n$ and $M=(M_p+M_n)/2\approx939$ MeV. 
All the form factors above are functions of $-(p_i-p_f)^2$. We may also define the isospin combinations $F_{1,2}^S=F_{1,2}^p+F_{1,2}^n$ and $F_{1,2}^V=F_{1,2}^p-F_{1,2}^n$. The values of the vector and the axial charged weak form factors at zero momentum transfer define the ``bare'' vector and axial coupling constants: $F_1^W(0)=\mathring{g}_V$, $G_A(0)=\mathring{g}_A$, which represent the Fermi and Gamow-Teller matrix element in neutron $\beta$-decay respectively. In particular, $\mathring{g}_V=1$ from isospin symmetry, and the correction due to the strong isospin-breaking effects is negligible due to  the Behrends-Sirlin-Ademollo-Gatto theorem~\cite{Behrends:1960nf,Ademollo:1964sr}. On the other hand, $\mathring{g}_A$ is not protected by any exact symmetry. We do not include the isospin-breaking correction to $\mathring{g}_A$ separately because it is already included in the respective first-principles calculations.

The nucleon mass difference $\Delta=M_n-M_p\approx1.3$ MeV and the electron mass $m_e\approx0.511$ MeV are much smaller than $M$. Therefore, the tree-level amplitude of the decay process $n(p_n)\rightarrow p(p_p)e(p_e)\bar{\nu}_e(p_\nu)$ is given by:
\begin{equation}
\mathcal{M}_\mathrm{tree}=-\frac{G_F}{\sqrt{2}}L_\lambda\bar{u}_{s'}(p)\gamma^\mu(\mathring{g}_V+\mathring{g}_A\gamma_5)u_s(p)+\mathcal{O}(\Delta^2)~,
\end{equation}
where $G_F=1.1663787(6)\times 10^{-5}$~GeV$^{-2}$ is the Fermi constant measured from the muon decay~\cite{Zyla:2020zbs}, $p=(p_n+p_p)/2$ is the average nucleon momentum, and  $L_\lambda=\bar{u}(p_e)\gamma_\lambda(1-\gamma_5)v(p_\nu)$ is the lepton piece. The recoil corrections scale as $\Delta/M\sim 10^{-3}$, which are small but important in precision physics. They were studied in detail with both conventional methods and effective field theory (EFT)~\cite{Holstein:1974zf,Wilkinson:1982hu,Ando:2004rk,Gudkov:2008pf,Ivanov:2012qe,Ivanov:2020ybx}, and will not be discussed here.

RCs of the order $\mathcal{O}(\alpha_{em}/\pi)$ must be included for a precise extraction of the weak coupling parameters. In the usual nomenclature, they are divided into the ``outer'' and ``inner'' corrections The former is a function of $\{E_e,m_e\}$ calculable within Quantum Electrodynamics (QED) and independent of details of strong interaction. The latter is instead a constant in $E_e$ but depends on details of the hadronic structure. The squared amplitude for the decay of a polarized neutron (to unpolarized final states) after the inclusion of the $\mathcal{O}(\alpha_{em}/\pi)$ RCs reads:
\begin{eqnarray}
|\mathcal{M}|^2&=&16G_F^2|V_{ud}|^2M_n M_p E_e(E_m-E_e)g_V^2(1+3\lambda^2)F(\beta)\left(1+\frac{\alpha_{em}}{2\pi}\delta^{(1)}\right)\left\{1+\left(1+\frac{\alpha_{em}}{2\pi}\delta^{(2)}\right)a_0\frac{\vec{p}_e\cdot\vec{p}_\nu}{E_eE_\nu}\right.\nonumber\\
&&\left.+\hat{e}_s\cdot\left[\left(1+\frac{\alpha_{em}}{2\pi}\delta^{(2)}\right)A_0\frac{\vec{p}_e}{E_e}+B_0\frac{\vec{p}_\nu}{E_\nu}\right]\right\}+\mathcal{O}(\Delta^3)~,
\end{eqnarray}
with \begin{equation}
a_0=\frac{1-\lambda^2}{1+3\lambda^2}~,\quad A_0=-\frac{2\lambda(\lambda+1)}{1+3\lambda^2}~,\quad B_0=\frac{2\lambda(\lambda-1)}{1+3\lambda^2}~.
\end{equation}
Here, $E_e=p_e\cdot p/M$ is the electron energy, $E_m=(M_n^2-M_p^2+m_e^2)/(2M_n)\approx M_n-M_p$ is the electron end-point energy, and $\beta=\sqrt{1-m_e^2M^2/(p_e\cdot p)^2}$ is the electron speed, all in the nucleon's rest frame. With these notations, the functions $\delta^{(1,2)}$ describe the outer corrections~\cite{Sirlin:1967zza,Garcia:1981it}:
\begin{eqnarray}
\delta^{(1)}&=&3\ln\frac{M_p}{m_e}-\frac{3}{4}+4\left(\frac{1}{\beta}\tanh^{-1}\beta-1\right)\left({\ln\frac{2(E_\mathrm{m}-E_e)}{m_e}}+\frac{E_m-E_e}{3E_e}-\frac{3}{2}\right)\nonumber\\
&&-\frac{4}{\beta}\mathrm{Li}_2\left(\frac{2\beta}{1+\beta}\right)+\frac{1}{\beta}\tanh^{-1}\beta\left(2+2\beta^2+\frac{(E_m-E_e)^2}{6E_e^2}-4\tanh^{-1}\beta\right)\nonumber\\
\delta^{(2)}&=&2\left(\frac{1-\beta^2}{\beta}\right)\tanh^{-1}\beta+\frac{4(E_m-E_e)(1-\beta^2)}{3\beta^2E_e}\left(\frac{1}{\beta}\tanh^{-1}\beta-1\right)\nonumber\\
&&+\frac{(E_m-E_e)^2}{6\beta^2E_e^2}\left(\frac{1-\beta^2}{\beta}\tanh^{-1}\beta-1\right)~,\label{eq:outer}
\end{eqnarray}
whereas $F(\beta)\approx 1+\alpha_{em}\pi/\beta$ is the Fermi's function that incorporates the Coulomb interaction between the final-state proton and the electron~\cite{Fermi:1934hr}. The function $\delta^{(1)}$ is also known as Sirlin's function $g(E_e,E_m)$.

\begin{figure}
	\begin{centering}
		\includegraphics[scale=0.3]{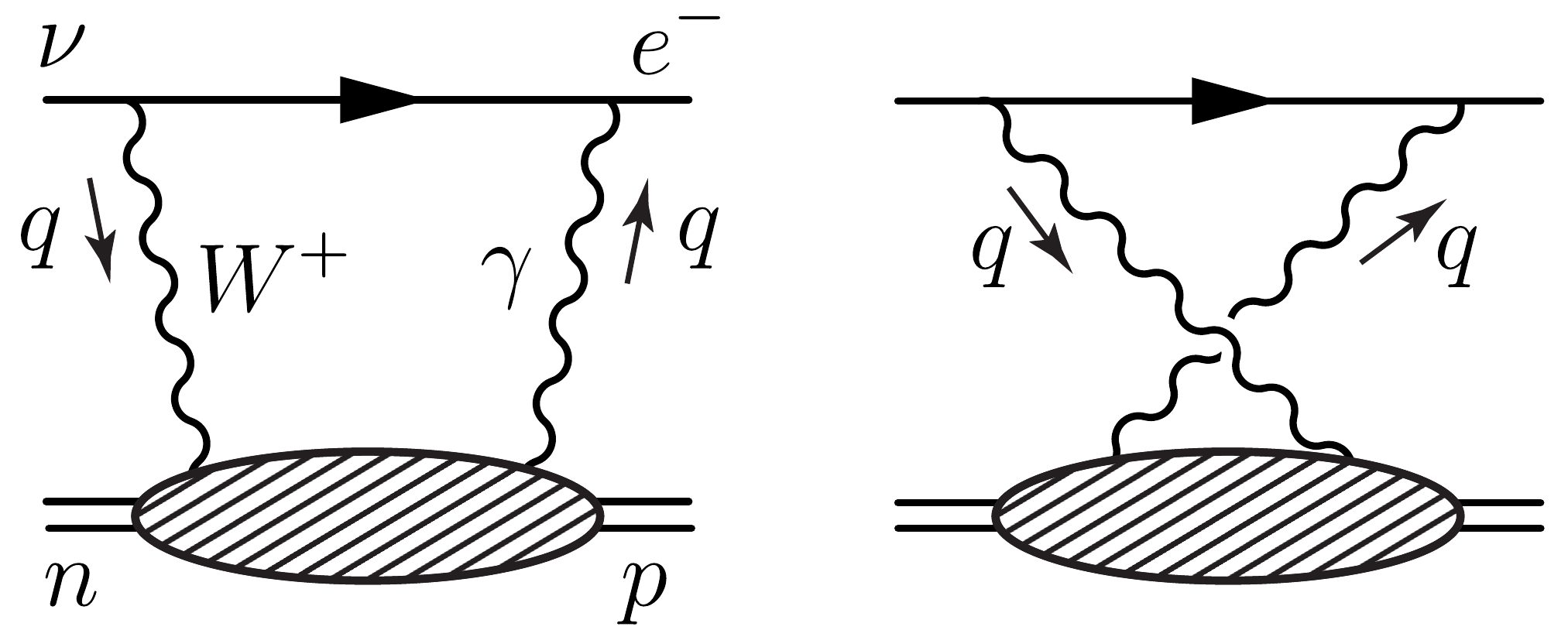}
		\hfill
		\par\end{centering}
	\caption{\label{fig:boxdiagram}The forward $\gamma W$-box diagrams in the free neutron $\beta$-decay.}
\end{figure}

The axial to vector coupling constants' ratio, parameter  $\lambda=g_A/g_V$, is understood as  fully renormalized by the inner RCs. In near-degenerate semileptonic $\beta$-decay processes, the inner RCs are most conveniently studied in Sirlin's representation~\cite{Sirlin:1977sv} (see also Refs.\cite{Seng:2019lxf,Feng:2020zdc,Seng:2020jtz} for a detailed account). In this formalism, most of the $\mathcal{O}(\alpha_{em}/\pi)$ electroweak RCs are either exactly known from current algebra, or give rise to the outer corrections in Eq.\eqref{eq:outer} and the Fermi's function. As a result, the renormalized vector and axial coupling constants read: 
\begin{eqnarray}
g_V&=&\mathring{g}_V\left\{1+\frac{\alpha_{em}}{4\pi}\left[3\ln\frac{M_Z}{M_p}+\ln\frac{M_Z}{M_W}+\tilde{a}_g\right]+\frac{1}{2}\delta_\mathrm{HO}^\mathrm{QED}+\Box_{\gamma W}^V\right\}\nonumber\\
g_A&=&\mathring{g}_A\left\{1+\frac{\alpha_{em}}{4\pi}\left[3\ln\frac{M_Z}{M_p}+\ln\frac{M_Z}{M_W}+\tilde{a}_g\right]+\frac{1}{2}\delta_\mathrm{HO}^\mathrm{QED}+\Box_{\gamma W}^A\right\}~,\label{eq:innerRC}
\end{eqnarray}
where $\tilde{a}_g$ is a pQCD correction factor and $\delta_\mathrm{HO}^\mathrm{QED}$ summarizes the leading-log higher-order QED effects~\cite{Marciano:1993sh,Erler:2002mv}. One observes that the fractional corrections to $\mathring{g}_V$ and $\mathring{g}_A$ are mostly identical and cancel in the ratio $g_A/g_V$. The only exceptions are the constants $\Box_{\gamma W}^V$ and $\Box_{\gamma W}^A$ that describe the inner RCs originated from the $\gamma W$-box diagrams (see Fig.\ref{fig:boxdiagram}), which are the focus of this paper\footnote{In the existing literature, e.g. Refs.\cite{Seng:2018yzq,Seng:2018qru,Feng:2020zdc}, the quantity $\Box_{\gamma W}^V$ was written as $\Box_{\gamma W}^{VA}$, where the superscript indicates that it involves the product of a vector current and an axial current. In this paper, the superscript carries a different meaning, namely which weak coupling constant (vector or axial) they are correcting.}.
With the above, we obtain:
\begin{equation}
\lambda=\frac{\mathring{g}_A}{\mathring{g}_V}\left[1+\Box_{\gamma W}^A-\Box_{\gamma W}^V\right]~.\label{eq:lambda}
\end{equation}

\section{$\gamma W$-box diagram}
\label{sec:box}

The $\gamma W$-box correction is of the natural size $\alpha_{em}/\pi\sim 10^{-3}$. Taking into account recoil corrections $\sim\Delta/M,\,m_e/M$ on top of the overall $\alpha_{em}/\pi$ factor would bring us to accounting for effects in the $10^{-6}$ range that exceed the precision goal by two or three orders of magnitude. This defines the level of the detalization that is needed in our analysis. We will consistently set $\Delta=m_e=0$ throughout the calculation below, as well as the proton recoil. This approximation also leads to the neglect of the pion pole due to the partially-conserved axial current (PCAC) hypothesis: the pion pole contribution, when contracted with the lepton tensor, results in lepton mass terms which, as stated are neglected. This precision level is supported by the fact that the lowest hadronic state is separated by the pion mass $\sim140$ MeV which is about hundred times larger than  $\Delta$. Notice however that this approximation may not be as safe for nuclear $\beta$-decay where the available $Q$-values may be as large as 15-20 MeV which are comparable to the energy level of nuclear excitations.

The part of the $\gamma W$-box diagram amplitude that contributes to the inner correction must involve an antisymmetric tensor that stems from the lepton spinor structure. It reads:
\begin{equation}
\mathcal{M}_{\gamma W}^{\mathrm{inner}}=-\frac{G_F}{\sqrt{2}}L_\lambda I_{\gamma W}^\lambda~,
\end{equation}
where
\begin{equation}
I_{\gamma W}^\lambda=\bar{u}_{s}(p)\gamma^\lambda\left[\mathring{g}_V\Box_{\gamma W}^V+\mathring{g}_A\Box_{\gamma W}^A\gamma_5\right]u_s(p)=ie^2\int\frac{d^4q}{(2\pi)^4}\frac{M_W^2}{M_W^2-q^2}\frac{\epsilon^{\mu\nu\alpha\lambda}q_\alpha}{(q^2)^2}T_{\mu\nu}^{\gamma W}
\end{equation}
with $\epsilon^{0123}=-1$ in our convention. The forward generalized Compton tensor describing the $W^+n\rightarrow\gamma p$ process, is defined as:
\beqn
T^{\mu\nu}_{\gamma W}=\int dxe^{iqx}\langle p|T[J^\mu_{em}(x)J^{\nu}_W(0)]|n\rangle.\label{eq:IgammaW}
\eeqn

To extract $\Box_{\gamma W}^{V}$ and  $\Box_{\gamma W}^{A}$, we use following identities:
\begin{equation}
\frac{1}{2}\bar u_s(p)\gamma^\mu u_s(p)=p^\mu~,\:\:\frac{1}{2}\bar u_s(p)\gamma^\mu\gamma_5 u_s(p)=S^\mu~,
\end{equation}
where the spin vector $S^\mu$ is fixed by $S^2=-M^2$ and $S\cdot p=0$, we obtain:
\begin{eqnarray}
\Box_{\gamma W}^V&=&\frac{ie^2}{2M^2\mathring{g}_V}\int\frac{d^4q}{(2\pi)^4}\frac{M_W^2}{M_W^2-q^2}\frac{\epsilon^{\mu\nu\alpha\lambda}q_\alpha p_\lambda}{(q^2)^2}T_{\mu\nu}^{\gamma W}\nonumber\\
\Box_{\gamma W}^A&=&-\frac{ie^2}{2M^2\mathring{g}_A}\int\frac{d^4q}{(2\pi)^4}\frac{M_W^2}{M_W^2-q^2}\frac{\epsilon^{\mu\nu\alpha\lambda}q_\alpha S_\lambda}{(q^2)^2}T_{\mu\nu}^{\gamma W}~.\label{eq:intermediate}
\end{eqnarray}
In what follows, we use $\mathring{g}_A\approx \lambda=-1.2756$ as a normalization in the second expression. Since $\Box_{\gamma W}^A\sim 10^{-3}$, the error induced by the ambiguity of $\mathring{g}_A$ is well below our precision goal. 

Only those components in $T_{\gamma W}^{\mu\nu}$ that contain an antisymmetric tensor contribute to Eq.\eqref{eq:intermediate}. These are
\beqn
T^{\mu\nu}_{\gamma W}&=&-\frac{i\epsilon^{\mu\nu\alpha\beta}q_\alpha p_\beta}{2(p\cdot q)}T_3+\frac{i\epsilon^{\mu\nu\alpha\beta}q_\alpha}{(p\cdot q)}
\left[S_\beta S_1+\left(S_\beta-\frac{(S\cdot q)}{p\cdot q}p_\beta\right)S_2\right]+...\label{eq:amplitudes}
\eeqn
The spin-independent, parity-violating amplitude $T_3$ and spin-dependent, parity-conserving amplitudes $S_{1,2}$ are functions of two invariants, $\nu=(p\cdot q)/M$ and $Q^2=-q_\mu q^\mu$. Plugging Eq.\eqref{eq:amplitudes} into Eq.\eqref{eq:intermediate} gives:
\begin{eqnarray}
\Box_{\gamma W}^V&=&\frac{e^2}{2M\mathring{g}_V}\int\frac{d^4q}{(2\pi)^4}\frac{M_W^2}{M_W^2+Q^2}\frac{1}{(Q^2)^2}\frac{\nu^2+Q^2}{\nu}T_3\nonumber\\
\Box_{\gamma W}^A &=&\frac{e^2}{M\mathring{g}_A}\int\frac{d^4q}{(2\pi)^4}\frac{M_W^2}{M_W^2+Q^2}\frac{1}{(Q^2)^2}\left\{\frac{\nu^2-{2Q^2}}{3\nu}{S_1}-\frac{Q^2}{\nu}{S_2}\right\}~,\label{eq:gVgA}
\end{eqnarray} 
where we have used the following identities:
\begin{eqnarray}
\int\frac{d^4q}{(2\pi)^4}q_\alpha F(\nu,Q^2)&=&\int\frac{d^4q}{(2\pi)^4}\frac{\nu}{M}p_\alpha F(\nu,Q^2)\nonumber\\
\int\frac{d^4q}{(2\pi)^4}q_\alpha q_\beta F(\nu,Q^2)&=&\int\frac{d^4q}{(2\pi)^4}\left[-\frac{\nu^2+Q^2}{3}g_{\alpha\beta}+\frac{4\nu^2+Q^2}{3}\frac{p_\alpha p_\beta}{M^2}\right]F(\nu,Q^2)\label{eq:symmetric}
\end{eqnarray}
that hold for any Lorentz scalar function $F(\nu,Q^2)$.

To evaluate the loop integrals, we need to discuss the symmetry properties of the Compton amplitudes. We start by considering the isospin structure of amplitudes $T_i,\,S_i$. Electromagnetic interaction does not conserve isospin and contains both isoscalar ($I=0$) and isovector ($I=1$) components. Therefore, each amplitude $A=T_i,\,S_i$ can be decomposed into components contributed by the isoscalar and isovector electromagnetic current respectively:
\beqn
A=A^{(0)}+A^{(1)}.\label{eq:decompose}
\eeqn
The two isospin amplitudes have a different behavior under $\nu\rightarrow-\nu$:
\beqn
A_i^{(I)}(-\nu,Q^2)=\xi_i^{(I)}A_i^{(I)}(\nu,Q^2),
\eeqn
with $\xi_i^{(I)}=\pm1$ and $\xi_i^{(0)}=-\xi_i^{(1)}$. It is easy to show that $\xi_i^{(0)}=-1$ for $T_3$ and $S_{1,2}$, so only the $I=0$ component of these amplitudes survives in the integrals in Eq.\eqref{eq:gVgA}.

\section{Dispersion representation of the forward Compton amplitudes}
\label{sec:DR}

\begin{figure}
	\begin{centering}
		\includegraphics[scale=0.5]{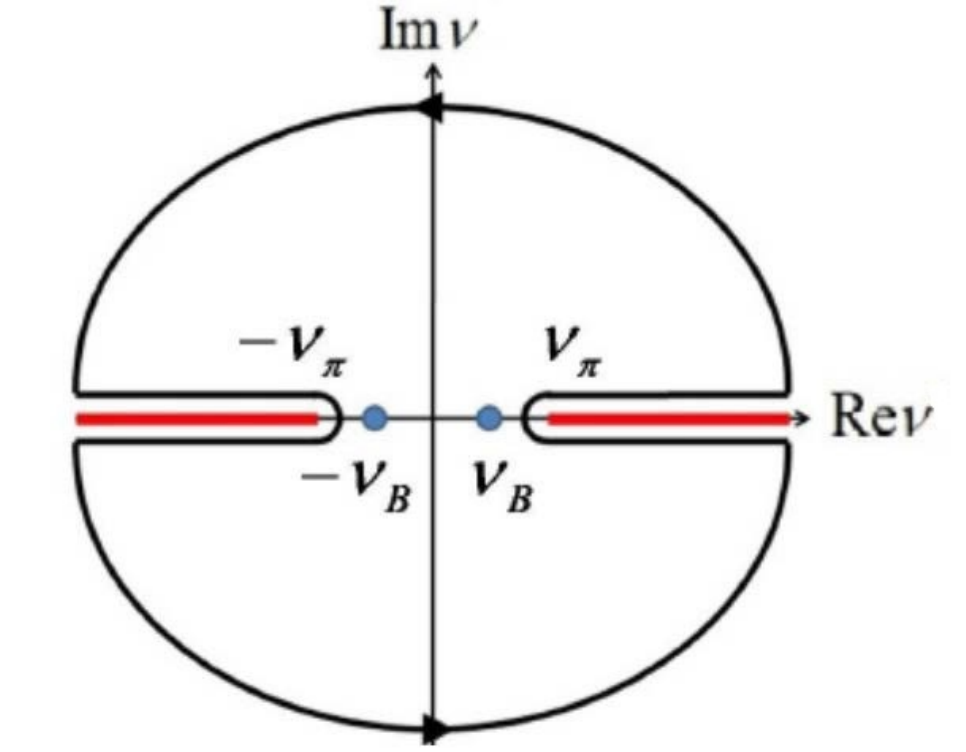}
		\hfill
		\par\end{centering}
	\caption{\label{fig:contour}The singularities of the forward Compton amplitudes on the complex-$\nu$ plane, and the contour chosen to derive the DRs in Eq.\eqref{eq:DR}.}
\end{figure}

Forward Compton amplitudes have singularities along the real axis $\nu$: poles due to a single nucleon intermediate state in the $s-$ and $u$-channels at $\nu=\pm \nu_B=\pm Q^2/(2M)$, respectively, and unitarity cuts at $\nu\geq \nu_\pi$ and $\nu\leq-\nu_\pi$ where {$\nu_\pi=(2Mm_\pi+m_\pi^2+Q^2)/(2M)$} is the pion production threshold (see Fig.\ref{fig:contour}). The discontinuity of the forward Compton tensor in the $s-$channel (i.e. $\nu\geq\nu_B$) is given by the generalization of the on-shell hadronic tensor to the $\gamma W$-interference:
\begin{equation}
\mathrm{Disc}T^{\gamma W}_{\mu\nu}(\nu)\equiv T^{\gamma W}_{\mu\nu}(\nu+i\varepsilon)-T^{\gamma W}_{\mu\nu}(\nu-i\varepsilon)=4\pi W_{\mu\nu}^{\gamma W}~,
\end{equation}
where
\begin{eqnarray}
W_{\mu\nu}^{\gamma W}&=&\frac{1}{4\pi}\sum_X(2\pi)^4\delta^{(4)}(p+q-p_X)\langle p|J_\mu^{em}(0)|X\rangle \langle X|J_\nu^W(0)|n\rangle\nonumber\\
&=&{-\frac{i\epsilon_{\mu\nu\alpha\beta}q^\alpha p^\beta}{2(p\cdot q)}F_3+\frac{i\epsilon_{\mu\nu\alpha\beta}q^\alpha}{(p\cdot q)}
\left[S^\beta g_1+\left(S^\beta-\frac{(S\cdot q)}{p\cdot q}p^\beta\right)g_2\right]+...}\label{eq:Wmunu}
\end{eqnarray}
The structure functions $F_3$ and $g_{1,2}$ can be decomposed similarly to $I=0,1$ components just like Eq.\eqref{eq:decompose}.

According to the crossing behavior established earlier and noticing that they cannot diverge faster than $\nu$ when $\nu\rightarrow\infty$, the amplitudes entering Eq. (\ref{eq:gVgA}) have the following dispersion representation:
\begin{eqnarray}
T_3^{(0)}(\nu,Q^2)&=&-4i\nu\int_{0}^\infty d\nu'\frac{ F_3^{(0)}(\nu',Q^2)}{\nu'^2-\nu^2}\nonumber\\ S_1^{(0)}(\nu,Q^2)&=&-4i\nu\int_{0}^\infty d\nu'\frac{ g_1^{(0)}(\nu',Q^2)}{\nu'^2-\nu^2}\nonumber\\
S_2^{(0)}(\nu,Q^2)&=&-4i\nu\int_{0}^\infty d\nu'\frac{ g_2^{(0)}(\nu',Q^2)}{\nu'^2-\nu^2}=-4i\nu^3\int_{0}^\infty d\nu'\frac{ g_2^{(0)}(\nu',Q^2)}{\nu'^2(\nu'^2-\nu^2)}~,\label{eq:DR}
\end{eqnarray}
where the $\nu$-integration is extended down to 0 to include the Born contribution. Notice that in the last line we have slightly modified the DR of $S_2^{(0)}$ using the Burkhardt-Cottingham (BC) sum rule~\cite{Burkhardt:1970ti}:
\begin{equation}
\int_0^1dxg_2(x,Q^2)=0~,
\end{equation}
where $x=Q^2/(2M\nu')$ is the Bjorken variable. This sum rule is a superconvergence relation and is expected to hold at all $Q^2$. The benefit of this treatment will become apparent in the later section. Substituting Eq.\eqref{eq:DR} into Eq.\eqref{eq:gVgA} and using the following Wick rotation formula~\cite{Marciano:1974tv},
\begin{equation}
\int\frac{d^4q}{(2\pi)^4}F(\nu,Q^2)=\frac{i}{8\pi^3}\int_0^\infty dQ^2 Q^2\int_{-1}^{+1}du\sqrt{1-u^2}F(iQu,Q^2)~,\label{eq:Wick}
\end{equation}
we can integrate the variable $u$ analytically to obtain our final dispersive representation of $\delta g_{V,A}^{\gamma W}$ as follows:
\begin{eqnarray}
\Box_{\gamma W}^V&=&\frac{\alpha_{em}}{\pi\mathring{g}_V}\int_0^\infty\frac{dQ^2}{Q^2}\frac{M_W^2}{M_W^2+Q^2}\int_0^1dx\frac{1+2r}{(1+r)^2}F_3^{(0)}(x,Q^2)\nonumber\\
\Box_{\gamma W}^A&=&-\frac{2\alpha_{em}}{\pi\mathring{g}_A}\int_0^\infty\frac{dQ^2}{Q^2}\frac{M_W^2}{M_W^2+Q^2}\int_0^1\frac{dx}{(1+r)^2}\left[\frac{5+4r}{3}g_1^{(0)}(x,Q^2)-\frac{4M^2x^2}{Q^2}g_2^{(0)}(x,Q^2)\right]~,
\end{eqnarray}
with $r=\sqrt{1+4M^2x^2/Q^2}$. As a useful crosscheck, the two-photon exchange correction to the hyperfine splitting in ordinary and muonic atoms is expressed through analogous two-fold integrals over electromagnetic spin structure functions~\cite{Carlson:2008ke,Carlson:2011af}.

The quantity $\Box_{\gamma W}^V$, relevant for the extraction of $V_{ud}$ from superallowed $\beta$-decays is well-studied within the dispersive approach~\cite{Seng:2018yzq,Seng:2018qru,Seng:2020wjq,Shiells:2020fqp} and is not addressed here. The main obstacle in those studies is the absence of direct experimental data of the structure function $F_3^{(0)}$. This forces one to either rely on data of $F_3$ from a different isospin channel (whose relation to $F_3^{(0)}$ contains a residual model-dependence), or from indirect lattice QCD data. On the other hand, despite having received much less attention, a high-precision dispersive analysis of $\Box_{\gamma W}^A$ is in fact much more robust because it depends on the parity-conserving, spin-dependent structure functions $g_{1,2}^{(0)}$. The isospin symmetry unambiguously relates them to $g_{1,2}^N$ measured in ordinary DIS:
\begin{equation}
g_{1,2}^{(0)}=\frac{1}{2}\left\{g_{1,2}^p-g_{1,2}^n\right\}~,\label{eq:isospin}
\end{equation}
the latter are defined via
\begin{eqnarray}
W_{\mu\nu}^{\gamma \gamma,N}&=&\frac{1}{4\pi}\sum_X(2\pi)^4\delta^{(4)}(p+q-p_X)\langle N|J_\mu^{em}(0)|X\rangle \langle X|J_\nu^{em}(0)|N\rangle\nonumber\\
&=&{\frac{i\epsilon_{\mu\nu\alpha\beta}q^\alpha}{(p\cdot q)}
\left[S^\beta g_1^N+\left(S^\beta-\frac{(S\cdot q)}{p\cdot q}p^\beta\right)g_2^N\right]+...}
\end{eqnarray}
Therefore, it is possible to perform a fully data-driven analysis of $\Box_{\gamma W}^A$ without introducing further model-dependence at low $Q^2$. We will perform such an analysis in the sections below. Following Sirlin's notation~\cite{Sirlin:1967zza}, we express our result as \begin{equation}\Box_{\gamma W}^A=\frac{\alpha_{em}}{2\pi}d=\frac{\alpha_{em}}{2\pi}\left[d_B+d_1+d_2\right]~,
\end{equation}
where $d_B$, $d_1$ and $d_2$ represent the elastic (Born) contribution, the inelastic contributions from $g_1^{(0)}$ and the inelastic contributions from $g_2^{(0)}$ respectively, which we will evaluate separately in the following sections.

\section{Elastic (Born) contribution}
\label{sec:born}

\begin{figure}
	\begin{centering}
		\includegraphics[scale=0.7]{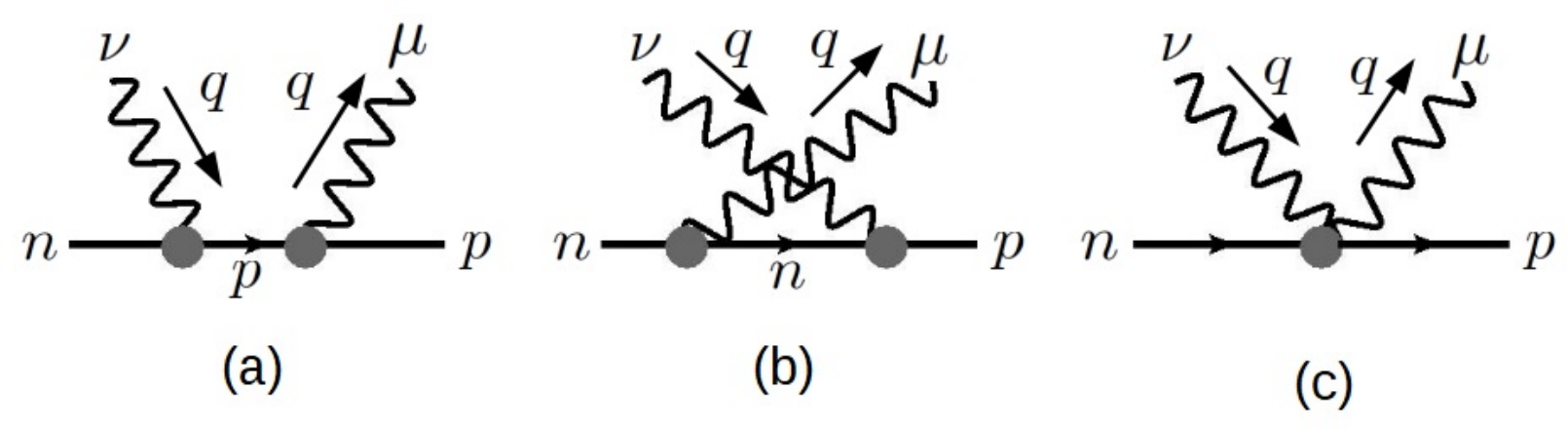}
		\hfill
		\par\end{centering}
	\caption{\label{fig:Born}The pole (a,b) and seagull (c) diagrams that contribute to $T^{\mu\nu}_{\gamma W}$.}
\end{figure}

Substituting $X=p$ into Eq.\eqref{eq:Wmunu} and using the elastic form factors defined in Eq.\eqref{eq:formfactors} give us the Born contribution to the spin structure functions needed for the evaluation of the box correction:
\beqn
g_1^{(0),B}=\frac{F_1^WG_M^S+F_1^SG_M^W}{8}\delta(1-x)~,\quad g_2^{(0),B}=-\tau\frac{F_2^WG_M^S+F_2^SG_M^W}{8}\delta(1-x)~,
\eeqn
where $G_E\equiv F_1-\tau F_2$ and $G_M\equiv F_1+F_2$ are the usual electric and magnetic Sachs form factors defined for both the electromagnetic and charged weak form factors, with $\tau=Q^2/(4M^2)$. All the form factors above are functions of $Q^2$ that drop at high $Q^2$, therefore one can neglect the $Q^2$ dependence of the $W$-boxon propagator $\sim Q^2/M_W^2$.
With this the Born contribution reads, 
\begin{equation}
d_B=-\frac{1}{2\mathring{g}_A}\int_0^\infty\frac{dQ^2}{Q^2}\frac{1}{(1+r_B)^2}\left\{\frac{5+4r_B}{3}\left[F_1^W G_M^S+F_1^S G_M^W\right]+\left[F_2^W G_M^S+F_2^S G_M^W\right]\right\}~,\label{eq:dB}
\end{equation}
where $r_B\equiv r|_{x=1}$.
We notice that $F_{1,2}^W=F_{1,2}^V$ by isospin symmetry, which means Eq.\eqref{eq:dB} is fully determined by the four nucleon electromagnetic form factors: $\{G_E^p,G_M^p,G_E^n,G_M^n\}$. Different parameterizations of these form factors~\cite{Drechsel:2002ar,Lorenz:2012tm,Lorenz:2014yda,Ye:2017gyb,Lin:2021umk,Lin:2021umz} all give consistent results within their respective error bars. In particular, the parametrization of Ref.~\cite{Ye:2017gyb} leads to
\begin{equation}
d_B=1.216(6)_{G_E^p}(9)_{G_M^p}(1)_{G_E^n}(2)_{G_M^n}=1.22(1)_\mathrm{FF}~.
\end{equation}
In particular, the central values of the contribution from $g_1^{(0),B}$ and $g_2^{(0),B}$ are 1.17 and 0.04 respectively. We observe that the latter is much smaller, which turns out to also be the case for the inelastic contributions. 

We pause here to comment on the Born contribution before moving on to the inelastic contributions. One may also try to derive it by calculating the Compton amplitudes from the first two Feynman diagrams in Fig.\ref{fig:Born} (the ``pole diagrams'') using the form factors in Eq.\eqref{eq:formfactors} as effective vertex functions, and then plugging them into Eq.\eqref{eq:gVgA}. The pole diagrams give:
\begin{eqnarray}
S_1^{(0),B}(\nu,Q^2)&=&-\frac{iM\nu Q^2}{Q^4-4M^2\nu^2-i\varepsilon}\left[F_1^W G_M^S+F_1^S G_M^W\right]+\frac{i\nu}{2M}F_2^W F_2^S\nonumber\\
S_2^{(0),B}(\nu,Q^2)&=&\frac{i\nu Q^4}{4M\left(Q^4-4M^2\nu^2-i\varepsilon\right)}\left[F_2^W G_M^S+G_M^W F_2^S\right]-\frac{i\nu}{4M}\left[F_2^W G_M^S+G_M^W F_2^S\right]~.\label{eq:BornCompton}
\end{eqnarray} 
We split each expression into two term, where the first term contains a singularity at $\nu^2=\nu_B^2$ and vanishes as $1/\nu$ when $\nu\rightarrow \infty$, while the second term is regular and diverges as $\nu$ when $\nu\rightarrow\infty$. It is easy to see that, retaining only the first term leads again to Eq.\eqref{eq:dB}, apart from a numerically small difference originating from our accounting for the BC sum rule in the DRs, effectively redefining the $g_2$ contribution to $d_B$. 
The regular terms in Eq.\eqref{eq:BornCompton} lead to an extra small deviation from Eq.\eqref{eq:dB}. The origin of this deviation lies in the Gerasimov-Drell-Hearn sum rule~\cite{Gerasimov:1965et,Drell:1966jv} and its extension to finite $Q^2$~\cite{Deur:2018roz} which relate the regular low-energy term to an integral over the inelastic part of $g_1$. 

This discussion simply means that the definition of the ``elastic'' contribution is not exactly the same in the diagrammatic and the dispersive representation. Of course, if we were able to calculate the full (i.e. pole + seagull) $T_{\gamma W}^{\mu\nu}$ exactly at all values of $\{\nu,Q^2\}$ with the diagrammatic approach, then the outcome must be identical to the DR analysis. But since this is impossible, the dispersive representation provides a much better starting point. We want to also point out that Refs.\cite{Hayen:2020cxh,Hayen:2021iga} attempted to calculate the Born contribution from the pole diagrams in Fig.\ref{fig:Born} (let us call it $d_B'$). 
\footnote{In the earlier versions of these Refs., the author made some algebraic mistakes when dealing with the symmetric loop integral of the form $\int d^4q q_\alpha q_\beta F(\nu,Q^2)$ (i.e. Eq.\eqref{eq:symmetric}). As a consequence, an incorrect analytic formula which gave an unexpectedly large value of $d_B'=2.64(3)$ was obtained. The published version of Ref.\cite{Hayen:2020cxh} corrected these mistakes, but retained only the term proportional to $(5+4r_B)/3$.} Should all terms be retained, they would have obtained the following result:
\begin{eqnarray}
d_B'&=&-\frac{1}{2\mathring{g}_A}\int_0^\infty\frac{dQ^2}{Q^2}\frac{1}{(1+r_B)^2}\biggl\{\frac{5+4r_B}{3}\left[F_1^W G_M^S+F_1^S G_M^W\right]+\left[F_2^W G_M^S+F_2^S G_M^W\right]-\frac{3\tau}{2}(1+r_B)^2F_2^W F_2^S\biggr\}\nonumber\\
&=&1.23(1)~\label{eq:dBCompton}
\end{eqnarray}
which only differs from the DR's definition of $d_B$ in Eq.\eqref{eq:dB} by a numerically small term. For the benefit of interested readers, we also provide in Appendix~\ref{sec:alternative} an alternative derivation of Eq.\eqref{eq:dBCompton} without making use of the invariant amplitudes.


\section{\label{sec:inelastic}Inelastic contributions}

Extensive measurements of the structure function $g_1^N$ were carried out in SLAC~\cite{Anthony:1993uf,Abe:1994cp,Abe:1997cx}, CERN~\cite{Adams:1994zd,Alexakhin:2006oza,Alekseev:2010hc,Aghasyan:2017vck}, DESY~\cite{Ackerstaff:1997ws} and JLab~\cite{Deur:2004ti,Wesselmann:2006mw,Deur:2008ej,Guler:2015hsw,Fersch:2017qrq}. In particular, we utilize the results from the EG1b experiment at JLab that measured the $g_1^p$~\cite{Fersch:2017qrq} and $g_1^n$~\cite{Guler:2015hsw} in a wide range of $\{x,Q^2\}$, from which the moments $\Gamma_i^N$ ($i=1,3,5$) were computed in bins of $Q^2$, common for $p$ and $n$, from 0.05~GeV$^2$ to 3.5~GeV$^2$. Full results are available in the supplementary material of each respective paper\footnote{There is a more recent measurement of $g_1^p$ from JLab at low $Q^2$~\cite{Zheng:2021yrn}, but unfortunately it does not measure $g_1^n$ simultaneously.}. 
Data on $g_2$ are generally scarce~\cite{Anthony:1999py,Anthony:2002hy,Amarian:2003jy,Wesselmann:2006mw,Kramer:2005qe,Fersch:2017qrq} and insufficient for a fully data-based analysis. Fortunately, its contribution is generally expected to be small due to the BC sum rule that forces the first moment of $g_2$ to vanish identically when accounting for elastic and inelastic contributions.

Since the data do not extend to an arbitrarily large $Q^2$ needed to evaluate the integrals, we make use of perturbative QCD results which are well under control theoretically above a separation scale $Q_0^2$, while directly using the experimental data that contains both perturbative and nonperturbative physics below that scale. 
Following our earlier works on the vector RC~\cite{Seng:2018yzq,Seng:2018qru} we take $Q_0^2=2$~GeV$^2$. For the vector RC case, not only does $Q_0^2=2$~GeV$^2$ mark the onset of pQCD regime, it also corresponds to the scale, below which the quality of data deteriorates severely leading to some sensitivity to $Q_0^2$. In the case of the axial RC, this scale lies well within the region covered by data, hence shifting it to a slightly higher value (not lower because pQCD description starts to break down) does not change the result.

\subsection{Contribution of $g_1^{(0)}$}

As shown in Eq.\eqref{eq:isospin}, the polarized structure function $g_1^{(0)}$ is simply related to $\{g_1^N\}$ that are measurable in DIS experiments. We define their moments as:
\begin{equation}
\Gamma_i^N(Q^2)\equiv\int_0^{x_\pi}x^{i-1}g_1^N(x,Q^2)~,\label{eq:moments}
\end{equation}
where $x_\pi=Q^2/\left[(M+m_\pi)^2-M^2+Q^2\right]$ is the pion production threshold. Notice that the definition above excludes the elastic contribution at $x=1$, which is a general convention adopted by most of the experimental papers. The first moment $\Gamma_1^{p-n}\equiv \Gamma_1^p-\Gamma_1^n$ is of particular interest because it satisfies the polarized Bjorken sum rule~\cite{Bjorken:1966jh,Bjorken:1969mm} at $Q^2\rightarrow \infty$. However, at large but finite $Q^2$ it receives a number of corrections~\cite{Ji:1993sv}:
\begin{equation}
\Gamma_{1,\mathrm{th}}^{p-n}(Q^2)=\frac{|\mathring{g}_A|}{6}C_{\mathrm{Bj}}(Q^2)+\sum_{i=2}^\infty\frac{\mu_{2i}^{p-n}}{Q^{2i-2}}~,\quad\text{large $Q^2$}~\label{eq:Bjorken}
\end{equation}
here the subscript ``th'' denotes the theory prediction (at large $Q^2$). 
The first term at the right hand side is the Bjorken sum rule with a pQCD correction factor\footnote{Please be reminded that one should not consider again the running effect of the QED coupling constant in $\Box_{\gamma W}^V$ and $\Box_{\gamma W}^A$, because it is already contained in the factor $\delta_\mathrm{HO}^\mathrm{QED}$ in Eq.\eqref{eq:innerRC}. So, throughout this paper we always take $\alpha_{em}=7.2973525693(11)\times 10^{-3}$ as a constant.}, while the second term summarizes the higher-twist (HT) effects starting from twist-four. The pQCD correction factor is written as:
\begin{equation}
C_\mathrm{Bj}(Q^2)=1-\sum_{n=1}^\infty \tilde{c}_n\left(\frac{\alpha_s}{\pi}\right)^n~,
\end{equation}
where $\alpha_s$ is the running strong coupling constant in the $\overline{\mathrm{MS}}$ scheme, 
while the coefficients $\{\tilde{c}_n\}$ are calculated at present to $n=4$~\cite{Baikov:2010iw,Baikov:2010je}:
\begin{eqnarray}
\tilde{c}_1&=&1\nonumber\\
\tilde{c}_2&=&4.583-0.333n_f\nonumber\\
\tilde{c}_3&=&41.44-7.607n_f+0.177n_f^2\nonumber\\
\tilde{c}_4&=&479.4-123.4n_f+7.697n_f^2-0.1037n_f^3~,
\end{eqnarray}
with $n_f$ the number of active quark flavors, and we refer the reader to Refs.~\cite{Seng:2018yzq,Seng:2018qru} for full detail of the pQCD contribution and relevant discussions and references. In the meantime, only the twist-four term among all the HT corrections needs to be included for our precision goal. There are several recent determinations of the coefficient $\mu_4^{p-n}$~\cite{Deur:2014vea,Kotlorz:2017wpu,Ayala:2018ulm} that are largely consistent with each other. In this work we quote the value $\mu_4^{p-n}=(-0.047\pm 0.020)M^2$ in Ref.\cite{Kotlorz:2017wpu}. We find that at 2~GeV$^2$, the inclusion of the twist-four correction reduces the size of $\Gamma_{1,\mathrm{th}}^{p-n}$ by about 13\%, but its total contribution to $d_1$ through the integral at $Q^2>2$~GeV$^2$ is only about 1\%. Coming back to our problem, we write
\begin{equation}
d_1=-\frac{3}{2\mathring{g}_A}\int_0^\infty\frac{dQ^2}{Q^2}\frac{M_W^2}{M_W^2+Q^2}\bar{\Gamma}_1^{p-n}(Q^2)~,\quad\bar{\Gamma}_1^{p-n}(Q^2)\equiv\int_0^{x_\pi}dx\frac{4(5+4r)}{9(1+r)^2}\left\{g_1^p(x,Q^2)-g_1^n(x,Q^2)\right\}~.
\end{equation}
When $Q^2\rightarrow\infty$ the function $\bar{\Gamma}_1^{p-n}$ reduces to $\Gamma_1^{p-n}$, but at low $Q^2$ the two are not identical due to target mass corrections $\sim M^2/Q^2$ contained in the factor {$f(x,Q^2)=4(5+4r)/(9(1+r)^2)$}, where $r=\sqrt{1+4x^2M^2/Q^2}$	

\begin{figure}
	\begin{centering}
		\includegraphics[scale=0.6]{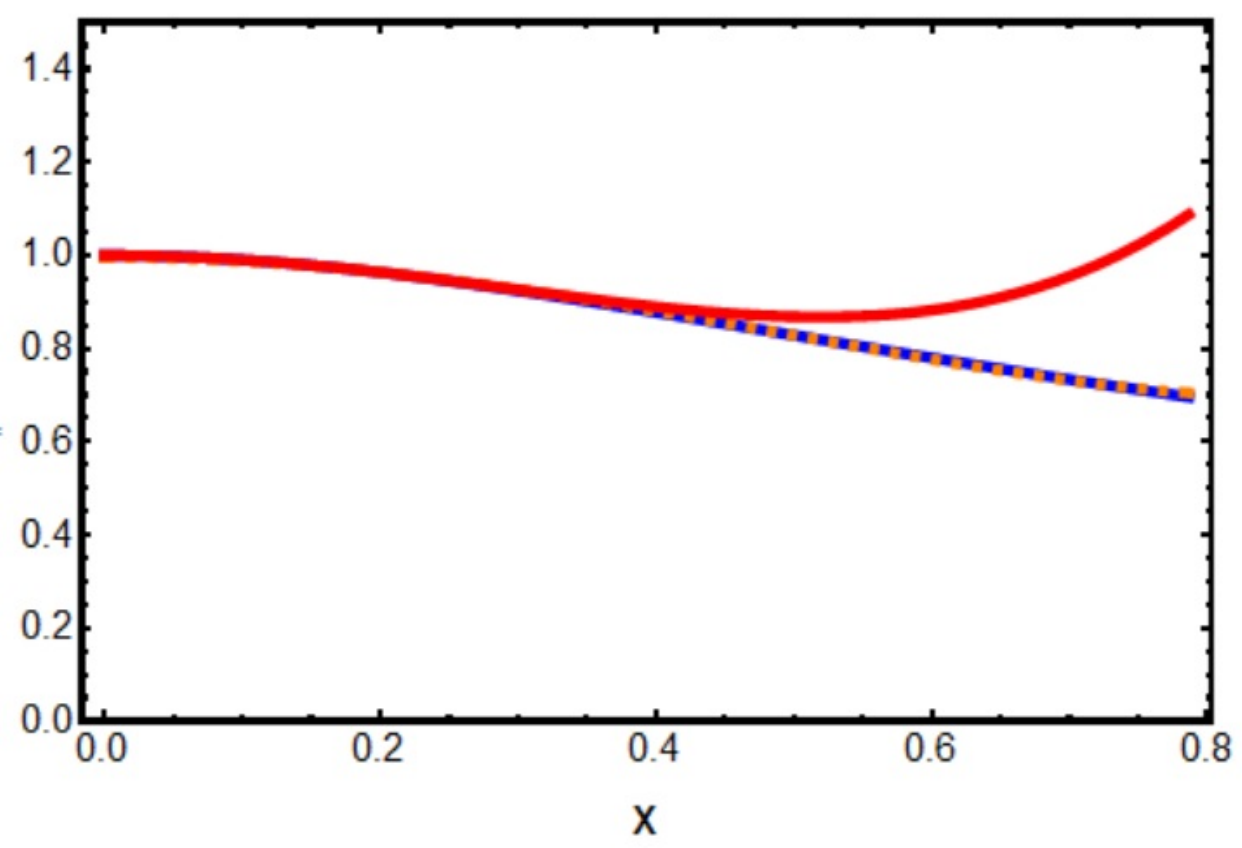}
		\includegraphics[scale=0.45]{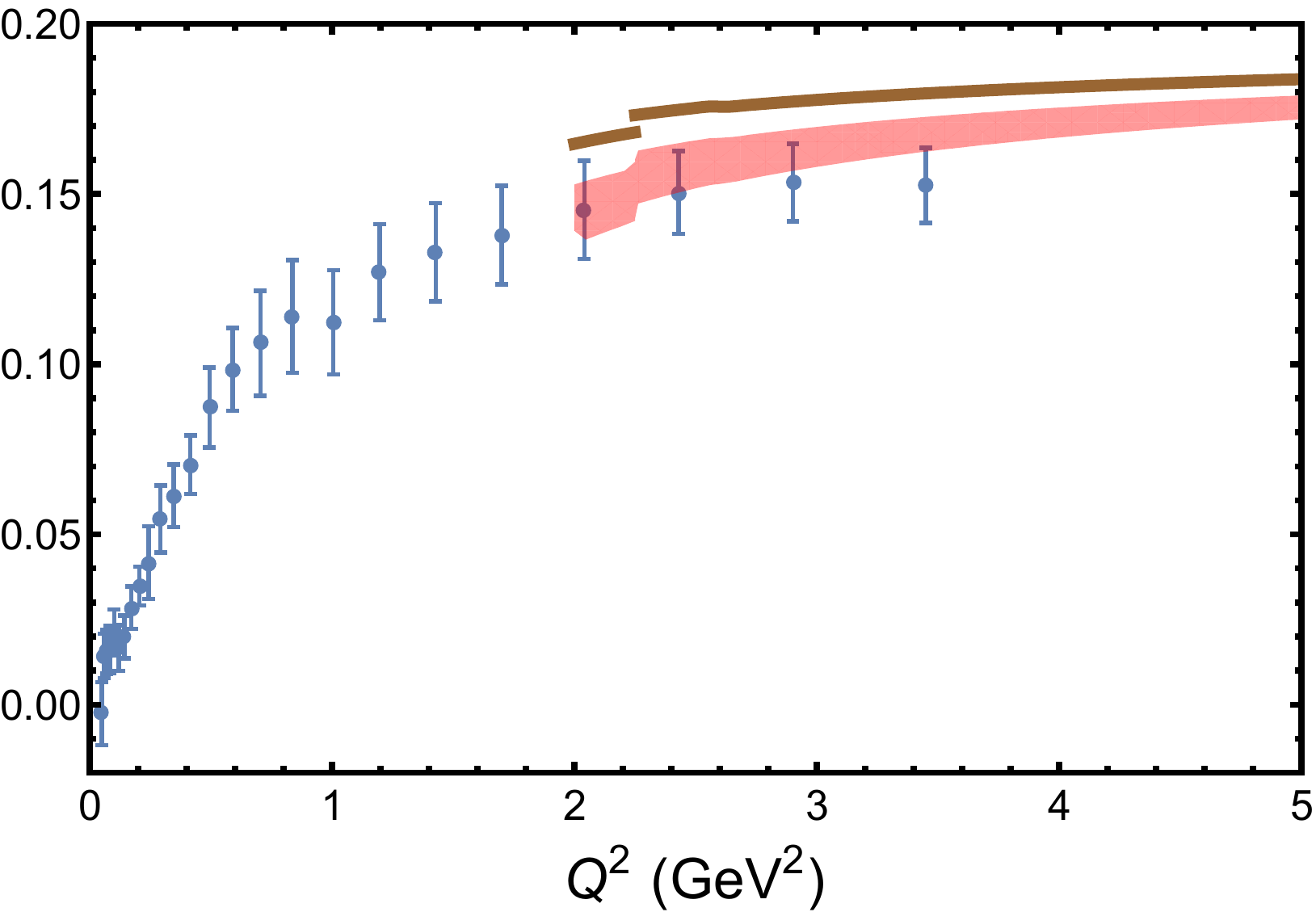}
		\hfill
		\par\end{centering}
	\caption{\label{fig:fit}Left panel: Comparison between the function $4(5+4r)/(9(1+r)^2)$ (blue solid curve), its global fit approximation (orange dashed curve) and the Taylor-expansion approximation (red solid curve) at $Q^2=1$~GeV$^2$, $0<x<x_\pi$. Right panel: Data points of $\bar{\Gamma}_1^{p-n}$ reconstructed from the EG1b experiment \cite{Fersch:2017qrq,Guler:2015hsw} versus the theory prediction at large $Q^2$ with (red band) and without (brown curve) the twist-four correction.}
\end{figure}

We use the following strategy to reconstruct the full $\bar{\Gamma}_1^{p-n}$ at low $Q^2$ from data. We fit  $f(x,Q^2)$ as function of $x$ at fixed $Q^2$ as
\begin{equation}
f(x,Q^2)= a(Q^2)+b(Q^2)x^2+c(Q^2)x^4+\dots,\quad 0<x<x_\pi~,\label{eq:globalfit}
\end{equation}
and obtain the three fitting parameters $a(Q^2)$, $b(Q^2)$ and $c(Q^2)$ by first dividing $0<x<x_\pi$ into, say, 1000 equal intervals, evaluating the 1000 respective discrete values for $f(x,Q^2)$, and performing a three-parameter fit with these discrete points using, e.g. \textit{Mathematica}. We find that this procedure allows for a very precise reproduction of the entire curve at $0<x<x_\pi$, where the difference between the original and the fitted curve is negligible for all practical purposes. Contrarily, a simple Taylor expansion in powers of $x^2M^2/Q^2$,
\begin{equation}
a(Q^2)+b(Q^2)x^2+c(Q^2)x^4\ncong 1-\frac{10M^2x^2}{9Q^2}+\frac{7M^4x^4}{3(Q^2)^2}\label{eq:Taylor}
\end{equation}
is only applicable at high $Q^2$ and low $x$ but significantly deviates for larger $x$. It breaks down completely at $Q^2=0$ where it becomes divergent, whereas $f(x,0)$ remains finite.
As an illustration, the two approximated expressions evaluated at a representative value of $Q^2=1$~GeV$^2$ are plotted together with the analytic form in the left panel of Fig.\ref{fig:fit}, and we clearly see that Eq.\eqref{eq:globalfit} nicely reproduces the latter for the full range of $x$. 

We thus reconstruct the full $\bar{\Gamma}_1^{p-n}$ in each bin of $Q^2$ in terms of the lowest Mellin moments,
\begin{equation}
\bar{\Gamma}_1^{p-n}(Q^2)= a(Q^2)\Gamma_1^{p-n}(Q^2)+b(Q^2)\Gamma_3^{p-n}(Q^2)+c(Q^2)\Gamma_5^{p-n}(Q^2)+\dots~,
\end{equation}
where $\Gamma_i^{p-n}$ ($i=1,3,5$) are obtained from Refs.\cite{Fersch:2017qrq,Guler:2015hsw}. We find that the difference between $\bar{\Gamma}_1^{p-n}$ and $\Gamma_1^{p-n}$ (i.e., the effect of higher moments) does not exceed 3.5\% for $Q^2=2~$GeV$^2$, which implies a negligible difference in the integral at $Q^2>2~$GeV$^2$. Therefore we will not distinguish between the two above 2~GeV$^2$. In contrast, the higher-twist correction due to $\mu_4^{p-n}$ reaches 13\% and needs to be kept along.

The right panel in Fig.\ref{fig:fit} shows the reconstructed $\bar{\Gamma}_1^{p-n}$ data points versus the large-$Q^2$ theory prediction using Eq.\eqref{eq:Bjorken}. We find that the theory and experiment match well at $Q^2>2$~GeV$^2$ (observe how the twist-four correction is needed to reconcile the two), which justifies our choice of $Q_0^2=2$~GeV$^2$ as the separation scale between the perturbative and non-perturbative regime. We therefore evaluate $d_1$ separately in these two regions. At $Q^2<Q_0^2$, we fit three curves that correspond to the upper bounds, central values and lower bounds of the discrete data points respectively, and evaluate the $Q^2$-integral and its uncertainty by integrating these three curves. Since the uncertainties of the data points are mainly systematics, this prescription takes into account the possible positive correlation effects. The resulting uncertainty is thus a conservative one; it is likely that it can further be reduced, but this would require a dedicated study of the systematic uncertainties of the data which lies beyond the scope of the present work. 
Meanwhile, at $Q^2>Q_0^2$ we evaluate the integral using the theory prediction in Eq.\eqref{eq:Bjorken}. The results are as follows:
\begin{eqnarray}
d_1^<&=&-\frac{3}{2\mathring{g}_A}\int_0^{Q_0^2}\frac{dQ^2}{Q^2}\frac{M_W^2}{M_W^2+Q^2}\left[\bar{\Gamma}_1^{p-n}(Q^2)\right]_\mathrm{data}=0.30(4)_\mathrm{data}\nonumber\\
d_1^>&=&-\frac{3}{2\mathring{g}_A}\int_{Q_0^2}^\infty\frac{dQ^2}{Q^2}\frac{M_W^2}{M_W^2+Q^2}\left[\frac{|\mathring{g}_A|}{6}C_{\mathrm{Bj}}(Q^2)+\frac{\mu_{4}^{p-n}}{{Q^2}}\right]=1.83(1)_\mathrm{HT}~,
\end{eqnarray}
where we neglected the uncertainty associated with the leading twist contribution compared to those coming from the data at low $Q^2$ and the HT correction (i.e. the coefficient $\mu_4^{p-n}$) at high $Q^2$. 
The total contribution of $g_1$ reads, 
\begin{equation}
d_1=d_1^<+d_1^>=2.14(4)_\mathrm{data}(1)_\mathrm{HT}~.
\end{equation} 
We note that the integral below $Q^2\leq0.05$ GeV$^2$  not covered by the data  but making part of $d_1^<$ is controlled by the isovector GDH sum rule~\cite{Gerasimov:1965et,Drell:1966jv}, $d\Gamma_1^{p-n}(Q^2=0)/dQ^2=(\kappa_n^2-\kappa_p^2)/8M^2$, with $\kappa_{p,n}$ denoting the proton's (neutron's) anomalous magnetic moment, respectively. Connecting the GDH-fixed value at $Q^2=0$ to the lowest data point produces a negligible $d_1^{\mathrm{low}~Q^2}\lesssim0.003$ contribution which is safely accommodated within the uncertainty. 

\subsection{Contribution of $g_2^{(0)}$}

In fact, we have already implemented this sum rule in the derivation of the DR of $S_2^{(0)}$. We emphasize the importance of this procedure for a reliable estimate of the contribution of $g_2^{(0)}$ to $\Box_{\gamma W}^A$: since experimental data typically only cover the inelastic region, enforcing an exact vanishing of the first moment of $g_2^{(0)}$ while operating with phenomenological parametrizations of different pieces can be a delicate matter. The explicit use of the BC sum rule thus precludes any numerically significant mistake caused by an imperfection of these parametrizations. 
As a result, the dispersive representation of $\Box_{\gamma W}^A$ only contains higher moments of $g_2^{(0)}$, in which the non-perturbative physics at small $x$ is suppressed. Additionally, since every extra power of $x^2$ is accompanied by $1/Q^2$, the contribution of $g_2^{(0)}$ bears no large logarithms.
Due to the smallness of the $g_2$ contribution, we opt for an approximate treatment and include this result in the estimate of the systematic uncertainty.

The inelastic contribution coming from $g_2^{(0)}$ reads,
\begin{eqnarray}
d_2&=&\frac{2}{\mathring{g}_A}\int_0^\infty\frac{dQ^2}{Q^2}\frac{M_W^2}{M_W^2+Q^2}\int_0^{x_\pi}\frac{dx}{(1+r)^2}\frac{4M^2x^2}{Q^2}\left[g_2^{p}(x,Q^2)-g_2^{n}(x,Q^2)\right]~,
\end{eqnarray}
where the isospin relation in Eq.\eqref{eq:isospin} is used. 
Rather than relying on data on $g_2^N$, we decompose $g_2$ into twist-two and twist-three (and higher) components~\cite{Zyla:2020zbs}, {$g_2^N=g_{2,\mathrm{tw}2}^N+g_{2,\mathrm{tw3}+}^N$}, and use the Wandzura-Wilczek relation~\cite{Wandzura:1977qf} for the former, 
\beqn
g_{2,\mathrm{tw}2}^N(x,Q^2)=-g_1^N(x,Q^2)+\int\limits_x^1\frac{dy}{y}g_1^N(y,Q^2)~.
\eeqn
Notice that the relation above should be understood to not contain the elastic contribution at $x=1$, because otherwise one could take $x_\pi<x<1$ at both sides, and then the left hand side and the first term at the right hand side would vanish but the second term at the right hand side would not, which is a contradiction. With this in mind, we multiply both sides by $x^{i-1}$ and integrate them at $0<x<x_\pi$ to obtain:
\begin{equation}
\int_0^{x_\pi}x^{n-1}g_{2,\mathrm{tw}2}^N(x,Q^2)=\frac{1-n}{n}\Gamma_n^N(Q^2)~,\label{eq:g2moments}
\end{equation}
where $\Gamma_n^N$ is defined in Eq.\eqref{eq:moments}. Therefore, we may use the available information of $g_1$ to evaluate the twist-two contribution to $d_2$. 

Again, we discuss the integral at large and small $Q^2$ separately. For $Q^2>Q_0^2$, we keep the leading term,
\begin{equation}
d_{2,\mathrm{tw}2}^>\approx-\frac{4}{3\mathring{g}_A}\int_{Q_0^2}^\infty \frac{dQ^2}{Q^2}\frac{M_W^2}{M_W^2+Q^2}\frac{M^2}{Q^2}\Gamma_3^{p-n}(Q^2)~,
\end{equation}
where we have used Eq.\eqref{eq:g2moments} and set $r\rightarrow 1$. The $Q^2$ dependence of the $W$ propagator can also be safely neglected as the integral converges. 
There is no simple sum rule for $\Gamma_3^{p-n}$ at large $Q^2$, but we may adopt a na\"{\i}ve valence quark picture that assumes each valence quark carries 1/3 of the nucleon's momentum. This gives:
\begin{equation}
g_1^{p-n}(x)\approx \frac{|\mathring{g}_A|}{6}\delta\left(x-1/3\right)~,\quad \text{large $Q^2$}
\end{equation}
which automatically reproduces the free polarized Bjorken sum rule. This na\"{\i}ve picture predicts $\Gamma_3^{p-n}\approx 0.11\Gamma_1^{p-n}$, which we may check against the experimental data: At $Q^2\approx 3.4$~GeV$^2$, Refs.\cite{Fersch:2017qrq,Guler:2015hsw} give $\Gamma_1^{p-n}\approx 0.1558$, $\Gamma_3^{p-n}\approx 0.0128$, i.e. $\Gamma_3^{p-n}\approx 0.08\Gamma_1^{p-n}$. So our na\"{\i}ve picture overestimates the size of $\Gamma_3^{p-n}$ by some 30\%, an acceptable uncertainty given our precision goal. With the above, we obtain:
\begin{equation}
d_{2,\mathrm{tw}2}^>\approx\frac{2}{81}\int_{Q_0^2}^\infty \frac{dQ^2}{Q^2}
\frac{M^2}{Q^2}=\frac{2M^2}{81Q_0^2}= 0.010(3)~.
\end{equation}

Next we turn to the small-$Q^2$ region where accounting for the leading twist may not be sufficient. 
As before, the leading twist contribution 
\begin{eqnarray}
d_{2,\mathrm{tw}2}^<&=&\frac{2}{\mathring{g}_A}\int_0^{Q_0^2}\frac{dQ^2}{Q^2}
\int_0^{x_\pi}\frac{dx}{(1+r)^2}\frac{4M^2x^2}{Q^2}\left[g_{2,\mathrm{tw}2}^{p}(x,Q^2)-g_{2,\mathrm{tw}2}^{n}(x,Q^2)\right]~,
\end{eqnarray}
is reconstructed by using the WW relation and the $x$-integral related to measured moments of $g_1$ by performing a two parameter fit of the kinematical function $f'(x,Q^2)$ for fixed $Q^2$ (since the first moment is removed by the BC sum rule, a two-parameter fit is already sufficient for our precision goal). 
\begin{equation}
f'(x,Q^2)=\frac{1}{(1+r)^2}\frac{4M^2x^2}{Q^2}= b'(Q^2)x^2+c'(Q^2)x^4+\dots~,\quad 0<x<x_\pi~.
\end{equation}
Using Eq.\eqref{eq:g2moments}, we obtain
\begin{equation}
\int_0^{x_\pi}\frac{dx}{(1+r)^2}\frac{4M^2x^2}{Q^2}\left[g_{2,\mathrm{tw}2}^{p}(x,Q^2)-g_{2,\mathrm{tw}2}^{n}(x,Q^2)\right]\approx -\frac{2}{3}b'(Q^2)\Gamma_3^{p-n}(Q^2)-\frac{4}{5}c'(Q^2)\Gamma_5^{p-n}(Q^2)~.
\end{equation}
With $\Gamma_i^{p-n}$ taken from Refs.\cite{Fersch:2017qrq,Guler:2015hsw}, this gives,
\begin{equation}
d_{2,\mathrm{tw}2}^<\approx-\frac{2}{\mathring{g}_A}\int_0^{Q_0^2}\frac{dQ^2}{Q^2}
\left[\frac{2}{3}b'(Q^2)\Gamma_3^{p-n}(Q^2)+\frac{4}{5}c'(Q^2)\Gamma_5^{p-n}(Q^2)\right]= 0.01(1)~,
\end{equation}
where we assigned a conservative 100\% uncertainty to the entire contribution.

To quantify higher twist contributions we recall the definition of the ``color polarizability"~\cite{Shuryak:1981pi,Jaffe:1989xx}
\begin{equation}
{\mathbf{d}_2(Q^2)=3\int_0^1dxx^2[g_2(x,Q^2)-g_{2,\mathrm{tw2}}(x,Q^2)],}
\end{equation}
of which we only consider the inelastic part {$\bar{\mathbf{d}}_2(Q^2)$} coming from the interval $0\leq x\leq x_\pi$~\cite{Alarcon:2020icz} since the elastic part is already taken into account.
In terms of this polarizability and neglecting higher moments, we obtain for the contribution of twist-three and higher,
\begin{eqnarray}
d_{2,\mathrm{tw}3+}^<&=&{\frac{2M^2}{3\mathring{g}_A}\int_0^{Q_0^2}\frac{dQ^2}{Q^4}[\bar{\mathbf{d}}_2^p(Q^2)-\bar{\mathbf{d}}_2^n(Q^2)]}
~.
\end{eqnarray}
For numerical estimates, we rely on the recent analysis of generalized spin polarizabilities of the nucleon in baryon chiral effective theory~\cite{Alarcon:2020icz}. 
We obtain, assigning a conservative 100\% uncertainty, 
\begin{eqnarray}
d_{2,\mathrm{tw}3+}^<&=&0.03(3)
~.
\end{eqnarray}

Combining the various pieces we finally arrive at
\begin{equation}
d_2=0.05(3)
\end{equation} 
as our estimate of the total inelastic contribution from $g_2^{(0)}$. We observe that it is two orders of magnitude smaller than $d_1$, following the same hierarchy as in $d_B$.

The inelastic contribution in our DR analysis then reads $d_1+d_2=2.19(4)_\mathrm{data}(1)_\mathrm{HT}(3)_{g_2}$. This is to be compared with $2.31(9)$ from Refs.~\cite{Hayen:2020cxh,Hayen:2021iga}, and we see that the two do not quite agree within error bars. While being identical in the large-$Q^2$ treatment, their estimation of the low-$Q^2$ contribution is largely model-based, raising questions about the reliability of the uncertainty. In contrast, in our treatment the low-$Q^2$ contribution is completely fixed by experimental data without any further assumption, apart from the small {$g_{2}$} correction for which only few assumption were made, which will become testable as soon as new, higher-quality low-$Q^2$ data for $g_2$ will become available. 

\section{Final discussions}
\label{sec:final}

Collecting all the results from Sec.\ref{sec:born} and \ref{sec:inelastic} gives:
\begin{equation}
\Box_{\gamma W}^A=3.96(1)_\mathrm{FF}(5)_\mathrm{data}(1)_\mathrm{HT}(3)_{g_2}\times 10^{-3}=3.96(6)\times 10^{-3}~,
\end{equation}
where the uncertainties come from the elastic form factors, the low-$Q^2$ $g_1$ data, the HT-correction to $g_1$ and $g_2$, respectively. We compare this to our recent update of $\Box_{\gamma W}^V$ using indirect lattice inputs: $\Box_{\gamma W}^V=3.83(11)\times 10^{-3}$~\cite{Seng:2020wjq}. These two numbers are very close to each other, and in fact their difference is consistent with zero:
\begin{equation}
\Box_{\gamma W}^A-\Box_{\gamma W}^V=0.13 (11)_V(6)_A\times 10^{-3}~.\label{eq:AminusV}
\end{equation}
Using Eqs.\eqref{eq:lambda}, \eqref{eq:AminusV} and the PDG average $\lambda=-1.2756(13)$~\cite{Zyla:2020zbs}, we obtain:
\begin{equation}
\mathring{g}_A=-1.2754(13)_\mathrm{exp}(2)_\mathrm{RC}~,
\end{equation}
which is consistent with the result from the current best lattice QCD determination.
Our result indicates that there is no practical distinction between $\lambda$ and $\mathring{g}_A$, unless the experimental precision of the former and the lattice precision of the latter have reached $3\times 10^{-4}$ or better. 

We wrap up with some discussions of the future prospects. Within the same DR framework, a much better precision is achieved for $\Box_{\gamma W}^A$  {than} for $\Box_{\gamma W}^V$ thanks to the existence of high-quality data of the structure function $g_1$ at $Q^2<2$~GeV$^2$. On the other hand, the precision of $\Box_{\gamma W}^V$ is limited by the low-quality data of the structure function $F_3$ from neutrino (antineutrino)-nucleus scattering experiments in the 80s~\cite{Bolognese:1982zd,Allasia:1985hw}. Better-quality data may come from the Deep Underground Neutrino Experiment (DUNE) in the next decade~\cite{Acciarri:2016crz,Alvarez-Ruso:2017oui}.

It was pointed out that a direct lattice QCD calculation of $\Box_{\gamma W}^V$ is a promising way to proceed at the present stage~\cite{Seng:2019plg}. Several exploratory calculations of mesonic $\gamma W$-box diagrams have shown great success~\cite{Feng:2020zdc,Ma:2021azh} and the same technology is directly applicable to nucleon. 
At present, no such direct calculation on the nucleon is available yet. A more involved comparison of the DR result for $\Box_{\gamma W}^V$ with the lattice computation of the respective quantity on the pion, amended with further phenomenological ingredients shows a nearly perfect agreement~\cite{Seng:2020wjq}. Even with this reassuring agreement, it is not unthinkable of that a direct lattice calculation could still disagree with the phenomenological, DR-based evaluation. 
Examples of such an unexpected disagreement are the pion-nucleon sigma term $\sigma_{\pi N}$ (see Ref.\cite{Aoki:2019cca} and references therein) and, more recently, the hadronic vacuum polarization contribution to $g_\mu-2$~\cite{Borsanyi:2020mff,Aoyama:2020ynm}. They show that even carefully-performed first-principles calculations or fully data-driven analysis may still contain unknown, previously unanticipated systematic effects which may seriously affect the implications of the corresponding precision experiments. Given these precedents, it is always useful to cross-check the lattice calculations with alternative methods.
Our new result of $\Box_{\gamma W}^A$ is perfectly up to this task as it is a solid phenomenological determination with the uncertainty very well under control. 

Further effort from the DR side should be dedicated to RCs to the GT strength in nuclear mirror decays where a recent study~\cite{Hayen:2019nic} revealed inconsistencies in the previous analyses. Removing these inconsistencies led to a better agreement for the $V_{ud}$ extracted across mirror and superallowed nuclear decays, as well as neutron decay. However, Ref.~\cite{Hayen:2019nic} only partially accounted for the $\gamma W$-box contribution. The dispersion formulation of the $\Box_{\gamma W}^A$ correction developed in this work can be directly applied to mirror systems. 
Following Refs.~\cite{Seng:2018qru,Gorchtein:2018fxl}, nuclear modifications of the universal free-neutron $\gamma W$-box correction can be computed, and we defer this task to future work. 



\section*{Acknowledgements} 

We are extremely thankful to Alexandre Deur and Xiaochao Zheng for their detailed explanations of the JLab experiments. We also appreciate Leendert Hayen for many inspiring discussions. We are furthermore grateful to Vadim Lensky and Vladimir Pascalutsa for providing their code for computing the generalized spin polarizabilities. This work is supported in
part by EU Horizon 2020 research and innovation programme, STRONG-2020 project
under grant agreement No 824093 and by the German-Mexican research collaboration Grant No. 278017 (CONACyT)
and No. SP 778/4-1 (DFG) (M.G), by
the Deutsche Forschungsgemeinschaft (DFG, German Research
Foundation) and the NSFC through the funds provided to the Sino-German Collaborative Research Center TRR110 “Symmetries and the Emergence of Structure in QCD” (DFG Project-ID 196253076 - TRR 110, NSFC Grant No. 12070131001) (C.Y.S).



\begin{appendix}

\section{Alternative derivation of Eq.(\ref{eq:dBCompton})}
\label{sec:alternative}

In this Appendix we outline a derivation of Eq.\eqref{eq:dBCompton}, namely the elastic contribution in the diagrammatic representation, directly from the tensor $T^{\gamma W}_{\mu\nu}$ without going through its invariant amplitudes. It involves some interesting tricks to deal with fermionic spinors and thus is worthwhile to be displayed for pedagogical purposes.

We start by computing $T^{\gamma W}_{\mu\nu}$ from the direct (D) and crossed (C) pole diagrams in Fig.\ref{fig:Born}. Using the elastic form factors as effective vertices, we obtain:
\begin{equation}
T_{\mu\nu}^{\gamma W,B}= \frac{\bar{u}_{s}(p)\Gamma_{\mu\nu}^{\mathrm{D}}u_{s}(p)}{2M\nu-Q^{2}}+\frac{\bar{u}_{s}(p)\Gamma_{\mu\nu}^{\mathrm{C}}u_{s}(p)}{-2M\nu-Q^{2}}\label{eq:Tmunu}
\end{equation}
where 
\begin{eqnarray}
\Gamma_{\mu\nu}^{\mathrm{D}} & = & \left[F_{1}^{p}\gamma_{\mu}-\frac{iF_{2}^{p}}{2M}\sigma_{\mu\alpha}q^{\alpha}\right]i\left[\slashed{p}+\slashed{q}+M\right]\left[F_1^W\gamma_{\nu}+\frac{iF_2^W}{2M}\sigma_{\nu\beta}q^{\beta}+G_{A}\gamma_{\nu}\gamma_{5}-\frac{G_{P}}{2M}\gamma_{5}q_{\nu}\right]\nonumber \\
\Gamma_{\mu\nu}^{\mathrm{C}} & = & \left[F_1^W\gamma_{\nu}+\frac{iF_2^W}{2M}\sigma_{\nu\beta}q^{\beta}+G_{A}\gamma_{\nu}\gamma_{5}-\frac{G_{P}}{2M}\gamma_{5}q_{\nu}\right]i\left[\slashed{p}-\slashed{q}+M\right]\left[F_{1}^{n}\gamma_{\mu}-\frac{iF_{2}^{n}}{2M}\sigma_{\mu\alpha}q^{\alpha}\right]
\end{eqnarray}
are matrices in the Dirac space.

We can get rid of the nucleon spinors $\bar{u}_{s}(p)$, $u_{s}(p)$
in the expression of $T_{\mu\nu}^{\gamma W,B}$ using the
following trick. First, we recall that any Dirac structure $\Gamma$
can be decomposed in terms of standard Dirac basis $1$, $\gamma_{5}$,
$\gamma^{\alpha}$, $\gamma^{\alpha}\gamma_{5}$, $\sigma^{\alpha\beta}$
using the following identity:
\begin{equation}
\Gamma=\frac{1}{4}\mathrm{Tr}[\Gamma]+\frac{1}{4}\mathrm{Tr}[\gamma_{5}\Gamma]\gamma_{5}+\frac{1}{4}\mathrm{Tr}[\gamma_{\alpha}\Gamma]\gamma^{\alpha}-\frac{1}{4}\mathrm{Tr}[\gamma_{\alpha}\gamma_{5}\Gamma]\gamma^{\alpha}\gamma_{5}+\frac{1}{8}\mathrm{Tr}[\sigma_{\alpha\beta}\Gamma]\sigma^{\alpha\beta}.\label{eq:Diracbasis}
\end{equation}
Next, we have the following identities when a Dirac basis is sandwiched between $\bar{u}_s(p)$ and $u_s(p)$:
\begin{eqnarray}
\bar{u}_{s}(p)u_{s}(p) & = & 2M\nonumber \\
\bar{u}_{s}(p)\gamma_{5}u_{s}(p) & = & 0\nonumber \\
\bar{u}_{s}(p)\gamma^{\alpha}u_{s}(p) & = & 2p^{\alpha}\nonumber \\
\bar{u}_{s}(p)\gamma^{\alpha}\gamma_{5}u_{s}(p) & = & 2S^{\alpha}\nonumber \\
\bar{u}_{s}(p)\sigma^{\alpha\beta}u_{s}(p) & = & \frac{2}{M}\epsilon^{\alpha\beta\rho\sigma}p_{\rho}S_{\sigma}~.\label{eq:sandwich}
\end{eqnarray}
Combining Eqs.\eqref{eq:Diracbasis} and \eqref{eq:sandwich}, we obtain:
\begin{equation}
\bar{u}_{s}(p)\Gamma_{\mu\nu}^{i}u_{s}(p)=\frac{M}{2}\mathrm{Tr}[\Gamma_{\mu\nu}^{i}]+\frac{p^{\alpha}}{2}\mathrm{Tr}[\gamma_{\alpha}\Gamma_{\mu\nu}^{i}]-\frac{S^{\alpha}}{2}\mathrm{Tr}[\gamma_{\alpha}\gamma_{5}\Gamma_{\mu\nu}^{i}]+\frac{1}{4M}\epsilon^{\alpha\beta\rho\sigma}p_{\rho}S_{\sigma}\mathrm{Tr}[\sigma_{\alpha\beta}\Gamma_{\mu\nu}^{i}],
\end{equation}
where $i=\mathrm{D,C}.$ The right hand side is free from the
nucleon spinors. The trace of the Dirac matrices can be performed using various
\textit{Mathematica} packages, so we do not display
the explicit results here.

We then plug our expression of $T_{\mu\nu}^{\gamma W,B}$, now free from nucleon spinors, into Eq.\eqref{eq:intermediate} and use Eq.\eqref{eq:symmetric} to simplify the integrals. Finally, we perform the Wick rotation using Eq.\eqref{eq:Wick} and integrate out the variable $u$ analytically. This brings us exactly to Eq.\eqref{eq:dBCompton}. 

\end{appendix}


\end{document}